# ReaxFF reactive force field study of polymerization of polymer matrix in carbon nanotube – composite system

Behzad Damirchi[a], Matthew Radue[b], Krishan Kanhaiya[c], Hendrik Heinz[c], Gregory Odegard[b], Adri C.T. van Duin[a*]

[a] *Department of Mechanical Engineering, Pennsylvania State University, University Park, PA 16802, USA*

[b] *Department of Mechanical Engineering, Michigan Technological University, Houghton, MI 49931, USA*

[c] *Department of Chemical and Biological Engineering, University of Colorado at Boulder, Boulder, CO 80309, USA*

[*] *Corresponding author. Department of Mechanical Engineering, Pennsylvania State University, University Park, PA, 16802, USA. E-mail address: acv13@psu.edu (A.C.T. van Duin).*

## Abstract

Human transport to Mars and deep space explorations demand the development of new materials with extraordinary high performance-to-mass ratios. Promising candidates to fulfill these requirements are ultra-high strength lightweight (UHSL) materials, which consist of polymer matrices fortified by pristine carbon nanotubes (CNTs). Previous investigations have showed that with an increase in CNT diameter, its preferred configuration changes from a circular form to a flattened shape that can be obtained in high pressure or tension conditions. The ReaxFF reactive force field can reveal detailed chemical interactions at the atomistic scale. To enable ReaxFF simulations on CNT/polymer interfaces, we trained force field parameters to capture the proper structure of flattened carbon nanotubes (flCNTs), i.e. dumbbell-like shape CNTs, against available polymer consistent force field – interface force field (PCFF-IFF) data which had good proximity to density functional theory (DFT) data. In this study we used accelerated ReaxFF molecular dynamics simulation using the optimized force field to study the polymerization of diglycidyl ether of bisphenol F (Bis F) and diethyltoluenediamine (DEDTA) molecules in vicinity of circular and flattened carbon nanotubes. Our results indicate that the flat regions of flCNT are more favorable spots for the polymers to settle compared to curved regions due to higher binding energies. Moreover, higher dimer generation around flCNT results in more effective coating of the nanotube which leads to higher load transfer in compared to circular CNT. According to our results there is a high alignment between polymers and nanotube surface which is due to strong π-π interactions of aromatic carbon rings in the polymers and nanotubes. These atomistic ReaxFF simulations indicate the capability of this method to simultaneously observe the polymerization of monomers along with their interactions with CNTs.

## 1. Introduction

Carbon nanotubes (CNTs) have been known as the strongest and stiffest manmade material to date. Moreover, their excellent properties can be enhanced if these materials can be homogenously embedded and dispersed into the light-weight materials offered by polymer engineering [1]. The application of light-weight materials is highly crucial in aeronautics and aerospace industry since their structural mass is directly proportional to the amount of fuel being consumed, thus, design of optimized materials with high performance-to-mass ratios is a significant necessity. The application of high strength carbon nanotube yarns [2] and fibers [3] has been growing due to advances in commercial availability of these materials. However, the fabrication and optimization of CNT fiber resin-forced composites still have shortcomings and need further detailed understanding [4].

Carbon nanotubes and polymer thermoset are on opposite ends of the spectrum of materials mechanical properties. The polymers thermoset's Young's modulus starts from low values of 2GPa [5] while the carbon nanotube is among the stiffest materials with Young's modulus of about 1000GPa [6]. Moreover, comparing the state-of-art carbon fibers with tensile strength, modulus and strain rate at failure of 6.9GPa, 324GPa and 2% with carbon nanotubes with the aforementioned modulus, and strain rate at failure of 20%, indicates that carbon nanotubes are promising candidates to build UHSL materials [7-9]. In order to make a promising CNT-composite material, there are a couple of factors that are needed to be checked and optimized. One of the main issues of such materials is nanotube dispersion into the polymer matrix. Several polymer matrices has been investigated as reinforcement, such as polyethylene [10], polyamides [11], epoxy [12, 13]. For the case of epoxy, it has been observed that the stiffer polymers lead to weaker adhesion and load transfer between nanotube and matrix [14, 15], however, in the case of covalent enhancements, where the nanotube is functionalized, this conclusion cannot be applied since the covalent bond will change the intrinsic properties of CNT [16].

Another factor which helps to improve the CNT-composite function is to utilize the nanotube graphitic assemblage of nanocomposites and CNT/carbon fiber hybrid composite. In order to achieve high mechanical performance, the paradigm must shift from traditional CNT-composite materials to highly concentrated CNT structures. This goal can be fulfilled by implementing flattened carbon nanotubes, enhancing the load transfer due to maximum contact surface provided between CNT/CNT and CNT/polymer systems. Moreover, the flattened carbon nanotube has reactive edges, providing suitable areas for polymer bindings in which the reactivity can be engineered via addition of functional groups to maximize mechanical load transfer [4]

Molecular dynamics simulations are providing detailed observation of the interactions at the atomic level which lead to better understanding of materials behavior and design of specific components for the required purposes. Among the existing empirical force field methods, reactive molecular dynamics simulations can reveal the reactions along with the conformation of the systems leading to more realistic properties monitoring due to breaking/forming of chemical bonds. The ReaxFF [17, 18] reactive force field has been applied to a wide range of applications [19] specifically investigating the mechanical properties of graphene layers and $sp^2$ $sp^3$ conversion of carbon atoms [20] leading to defunctionalization of polymers and subsequent binding to carbon nanotubes. Recently, a new method has been developed that enables study of the crosslinking of the polymers by accelerating the target reactions so that the transition state of the reactions can be observed during the simulations. This method overcomes the shortcomings of normal MD simulations which might not lead to proper observation of all the reactions from reactants to the products do to time-scale issues [21]. This so-called bond boosts MD simulation, applies certain restraints to sets of atoms to make the crosslinking happen, however, the reaction will be rejected if the energy barrier is higher than the applied energy boost value.

In the current paper, we optimized the force field carbon parameters to properly capture the flattened carbon nanotube structure using PCFF-IFF [22, 23] data. Using the optimized force field, we applied accelerated molecular dynamics simulation to investigate the polymerization of Bis F and DEDTA molecules in vicinity of circular and flattened carbon nanotubes. The binding energies and reaction details are compared for each case and the behavior of the systems are explained based on the binding energies and molecular components comparison. Moreover, we studied the alignment of polymer matrix around the circular nanotube while the polymerization is taking place, showing that this polymer selection leads to a highly aligned structure.

## 2. Methodology

### 2.1 Simulation details, ReaxFF reactive force field

ReaxFF is an empirical based potential which can provide the details of chemical reactions since it can capture the bond forming/breaking during the simulation using bond order concept [24, 25]. The significance of this concept is that the potential terms are smoothly varying, which leads to continuity of energy terms in bond formations and dissociations. The bond order term is calculated based on the distance of pairs of atoms and will be applied on all bonded interaction, i.e. bond, angle and torsion energy terms. The non-bonded interactions such as van der Waals and Coulomb will be updated for all pairs of atoms regardless of connectivity. In order to have continuity in non-bonded energy terms, a seventh order Taper function [17] is applied so that the first, second and third derivatives will go to zero at cutoff distance of 10 Å. Generally, ReaxFF implementation is based on the following energy equation:

$$E_{\text{system}} = E_{bond} + E_{over} + E_{under} + E_{lp} + E_{val} + E_{tor} + E_{vdWaals} + E_{coulomb} + E_{trip} \qquad (1)$$

where $E_{bond}, E_{over}, E_{under}, E_{lp}, E_{val}, E_{tor}, E_{vdWaals}$ and $E_{trip}$ are representing bond energy, over-coordination energy penalty, under-coordination stability, lone-pair energy, valence angle energy, torsion angle energy, van der Waals energy, Coulomb and triple bond stabilization energies, respectively. ReaxFF has been applied in a wide range of applications such as battery materials [26], combustion [18, 27], crystals [17], polymers [28-30] and, recently, ferroelectric materials [31, 32].

## 2. 2 Force field parameters optimization

The accuracy of the force field parameters is key to properly simulating a process since all the reactions and interactions are directly governed by the force field. Therefore, the first step in each simulation is to justify the force field being used. In this section, we describe the results from the force field parameters improvement to properly capture the behavior and shape of flattened carbon nanotube and later, we will use that improved force field to study the crosslinking of the polymers over carbon nanotubes.

Radially collapsed single-walled carbon nanotubes (SWCNTs) are recognized as closed-edge graphene nanoribbons [33]. There are regions of stability in carbon nanotubes; $D < D_{meta} = 2.2$ nm where $D$ is the diameter of carbon nanotube, represents the region where circular shape is energetically stable, $D_{meta} < D < D_{abs}$ where $D_{abs} = 5.1 nm$ and both deformed and circular shapes can be stabilized, and for $D > D_{abs}$ the deformed shape is energetically stable. Hasagawa et. al performed a series of DFT-D2 simulations on various deformed carbon nanotubes. They showed that deformed carbon nanotubes can acquire peanut-like and dumbbell-like shapes where for $D > 4.4 nm$ the dumbbell-like shape is more stable [33]. Figure 1 shows a flattened (30,30) armchair CNT in a dumbbell-like shape.

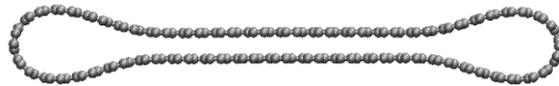

Figure 1. (30, 30) Dumbbell-like flattened carbon nanotube

To investigate the behavior of existing force fields on flattened carbon nanotubes, a series of energy minimization simulations was conducted using ReaxFF C-2013 [20], PCFF-IFF [22] and OPLS [34] and compared with DFT-D2 data [33]. For all candidate force fields, MD models were built of several armchair CNTs, namely, (20,20), (24,24), (30,30), (36,36), (40,40), (44,44), and (50,50) CNTs. Figure 2 illustrates the geometrical comparison of aforementioned force fields with DFT-D2 data set. The plotted flattened CNT width versus open CNT diameter indicates that among the existing force fields, PCFF-IFF can well reproduce the DFT-D2 data to predict the proper shape of the flattened CNT. OPLS underestimates the flattened CNT width, whereas ReaxFF overestimates it. ReaxFF C-2013 parameters cannot correctly maintain the curved region of flCNT and those areas are fully collapsed after energy minimization.

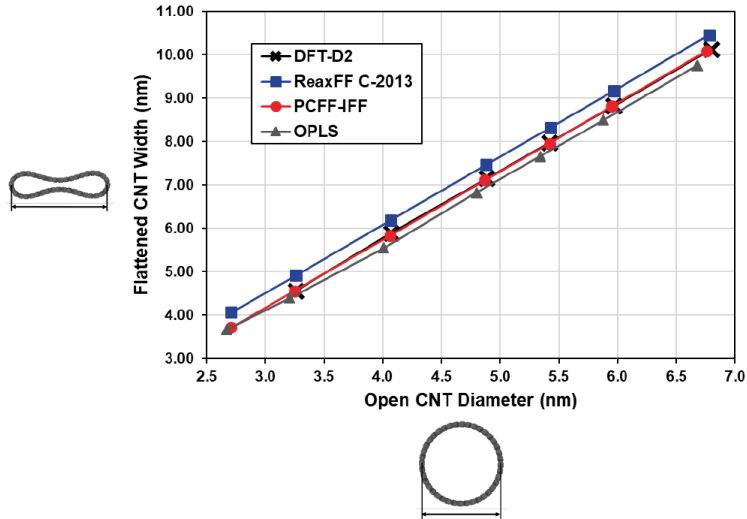

**Figure 2.** Flattened carbon nanotube geometrical comparison for DFT-D2, ReaxFF C-2013, PCFF-IFF and OPLS force fields.

Figure 3 depicts the energy comparison between existing force fields and DFT-D2 data set. In the diagram, the energy difference between flattened and open (circular) CNT is plotted versus the open CNT diameter. The positive values $of\ \Delta E = E_{flat} - E_{open}$ represent the region of stability where both circular and flattened shapes can be stabilized, however, for the negative values of $\Delta E$, flattened shape is energetically favorable. The diameter where $\Delta E = 0$ is known as threshold diameter. Based on the intersection of each curve with $\Delta E = 0$ axis, the threshold diameter estimated by OPLS acquires the highest value among other methods which is 6.2 nm. The high $\Delta E$ values predicted by OPLS can be attributed to low attraction between graphitic layers and/or high resistance to bending in graphene. PCFF-IFF slightly underestimates the relative flCNT energy but, relative to the other force fields analyzed, demonstrates good agreement with DFT-D2. Additionally, PCFF-IFF predicts the open CNT diameter of 4.2 nm and has the highest proximity to DFT-D2 with prediction of 5.1 nm. Based on the ReaxFF simulations, the calculated threshold dimeter value of 3.5 nm is underestimated. This suggests that the CNTs are too easily flattened in ReaxFF C-2013, which is consistent with the overly flattened predicted shape in Figure 4 (a). Accurate energies require an appropriate balance between the cohesive energy between graphitic layers, which captures the favorable energy of bringing the opposite sides of the CNT together, and the energy penalty associated with inducing high curvature at the bulb-like ends. Therefore, given that ReaxFF C-2013 shows low $\Delta E$ values, it is implied that either the attraction between graphitic layers is overly high, or the resistance to bending in graphene is too low, or a combination thereof. The results discussed above indicate that PCFF-IFF accurately reproduces the geometry and energy of flCNTs

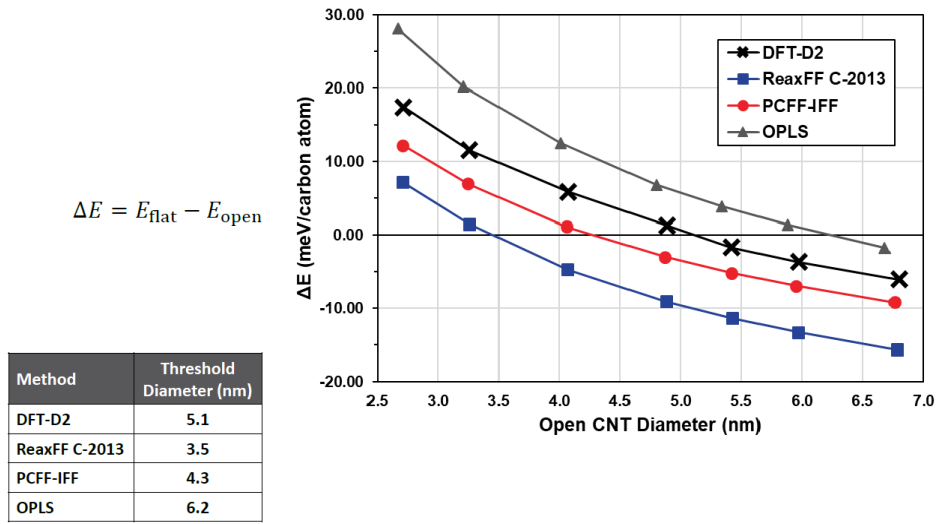

$\Delta E = E_\text{flat} - E_\text{open}$

| Method | Threshold Diameter (nm) |
|---|---|
| DFT-D2 | 5.1 |
| ReaxFF C-2013 | 3.5 |
| PCFF-IFF | 4.3 |
| OPLS | 6.2 |

**Figure 3.** Energy and threshold diameter comparison for DFT-D2, ReaxFF C-2013, PCFF-IFF and OPLS force fields.

To train the ReaxFF force field carbon parameters for flattened carbon nanotube, we used the PCFF-IFF data since we have straightforward access to the method and could simulate the various configurations of carbon nanotubes which will be discussed in the following. Thus, the energy difference between the flattened and circular carbon nanotubes are used as input to the ReaxFF trainset. However, after parameter optimization, we observed that the curved regions of flattened CNT could not maintain their curvature and collapse (Figure 4 (a)). To solve this problem, we used the PCFF-IFF force field to calculate the energy values of generated collapsed CNT and added it to the trainset. Figure 4 shows the nanotube modification to implement the PCFF-IFF. Figure 4 (a) is the output of the ReaxFF energy minimization, thus, in order to switch to the PCFF-IFF required model, a pair of dummy electrons are added to each carbon atom [22] and the atoms are restrained to keep the collapsed form during the simulation. Having the 3 sets of energy values for circular, flattened and collapsed carbon nanotube structure, the ReaxFF carbon parameters are re-optimized.

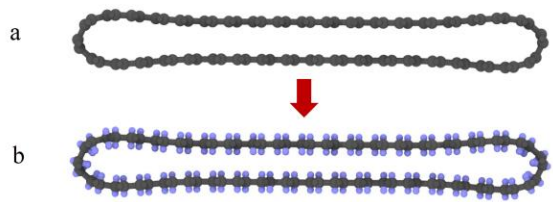

**Figure 4.** The collapsed structure of carbon nanotube after first ReaxFF parameter optimization (a). The PCFF-IFF input model with added dummy electron pairs.

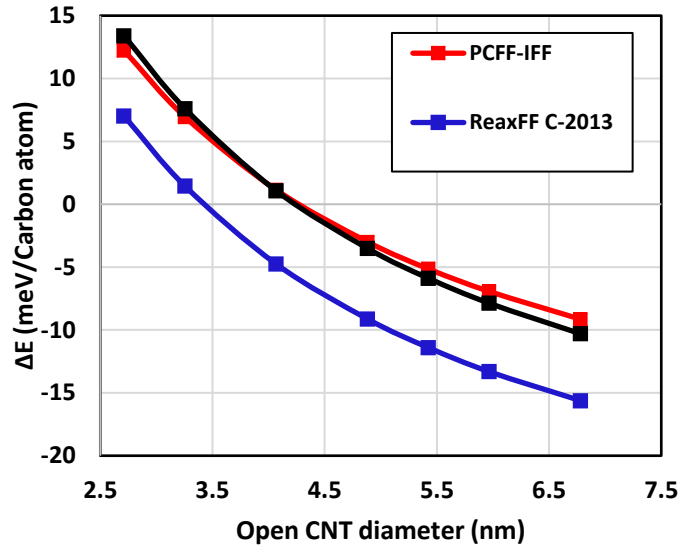

**Figure 5.** Comparison of ReaxFF C-2013 and optimized ReaxFF with PCFF-IFF energy values. The $\Delta E$ value is the energy difference between circular and flattened models.

As mentioned before, we used the ReaxFF C-2013 parameters as the initial values to train the force field. The carbon parameters are identical to the C/H/O by Chowdhury et al. [27], thus, we used the later force field, since it contains the optimized parameters for other atom types which we are going to use later to study the crosslinking of the polymers. Figure 5 illustrates the optimized ReaxFF force field and its comparison to the ReaxFF C-2013 and PCFF-IFF force fields. The optimized force field predicts the same 4.3 nm threshold diameter as PCFF-IFF. Therefore, the optimized force field has the carbon atom parameters refitted for proper flattened CNT shapes and energies along with the C/H/O/N parameters from the initial force field. We used this optimized force field to study the crosslinking of the DEDTA and Bis F molecules in vicinity of flattened and circular carbon nanotubes as will be discussed later in this article.

2. 3 Binding energy calculation

To validate our force field, the binding energies between DEDTA and Bis F molecules with circular are calculated and compared to the existing force fields. The binding energy between a monomer and carbon nanotube is defined as the energy difference between the two configurations that monomer and nanotube are adjacent versus apart from each other, i.e. nonadjacent. Binding energy can be calculated using equations (1-3):

$$B.E = (E_{CNT+Monomer})_{adjacent} - (E_{CNT} + E_{Monomer})_{nonadjacent} \tag{1}$$

$$E_{CNT+Monomer} = \frac{1}{n} \sum_{i=5:0.25fs:25:ps}^{n} E_{(CNT+Monomer)_i} \tag{2}$$

$$E_{CNT} + E_{Monomer} = \frac{1}{n} \sum_{i=5:0.25fs:25:ps}^{n} E_{(CNT)_i} + E_{(Monomer)_i} \tag{3}$$

where n is the number of time steps from 5ps to 25ps, i.e. 80000. The first term in right side of equation (1) is the energy of adjacent carbon nanotube and monomer, while the second term represents the nonadjacent configuration energy.

In order to justify the force field, a (10,10) circular carbon nanotube has been generated using VMD [35] and the binding energy between Bis F molecule and carbon nanotube is calculated using equations (1-3). Figure 6 illustrates the structures of adjacent and nonadjacent carbon nanotube and Bis F. Both structures are equilibrated for 25ps in NVT ensemble and the average energies from 5ps to 25ps are calculated. The absolute value of binding energy between circular nanotube and Bis F is obtained 18.74 kcal/mol which is in good agreement with the energy value of 23 kcal/mol reported by Gou et al. [36] using COMPASS [37] with a 6-9 Lenard-Jones non-bond interaction.

a b

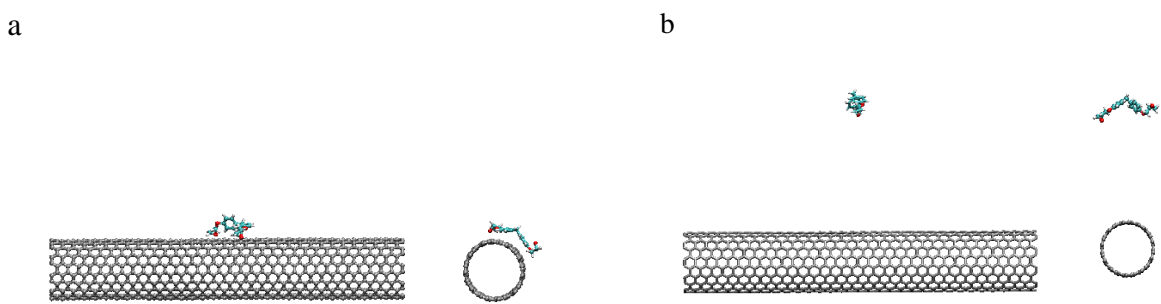

**Figure 6.** (a) The adjacent (b) nonadjacent configurations of carbon nanotube and Bis F

Applying the same approach, the binding energy of 23.46 kcal/mol is calculated between (15,15) circular CNT and Bis F. The obtained binding energies between DEDTA and flCNT in curved and flat parts are 15.15 kcal/mol and 24.57 kcal/mol, respectively (Figure 7). Based on the binding energies, it is observed that there is 9.42 kcal/mol energy difference between the curved and flat regions of flCNT with DEDTA molecule. According to the PCFF-IFF calculations, this energy difference is 9.50 kcal/mol, which has very good agreement with optimized ReaxFF force field. This energy difference will explain the behavior of flCNT/polymer system which will be discussed later.

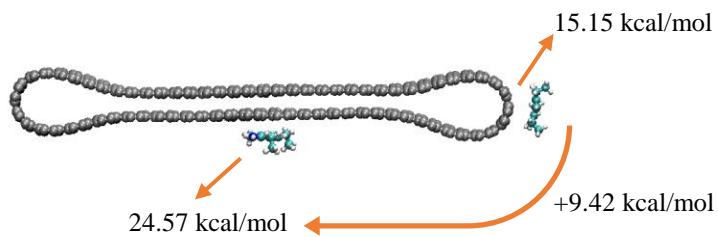

**Figure 7.** Binding energies between curved and flat parts of flCNT with DEDTA molecule

## 2. 4 Accelerated molecular dynamics

Recently, a method developed by Vashisth et al. [21] within the ReaxFF reactive force field framework has been used to accelerate the reaction by providing additional local energy to help the reactants overcome reaction barriers. The acceleration method basically adds a restraint energy to the system on the pairs of atoms which are within a defined distance from each other. Once it is activated, it will last for a predefined number of time steps – here we used 20,000 - focusing on the selected atoms while the rest of the system is behaving based on the general form of potentials in ReaxFF, i.e. equation 1. The polymerization reaction we want to observe in the current study is based on the amine-epoxy ring opening and proton transfer so that the DEDTA and Bis F molecules will be connected and generating dimers, trimers, etc. Figure 8 shows the structure of DEDTA and Bis F molecules.

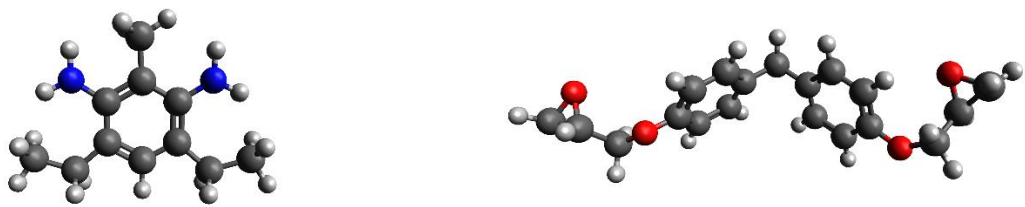

**Figure 8.** Molecular structures of DEDTA (left) and Bis F (right). The black, blue, red and white colors represent carbon, nitrogen, oxygen and hydrogen atoms, respectively.

## 2. 5 Polymerization and alignment of DEDTA and Bis F in vicinity of carbon nanotube

In order to study the polymerization of DEDTA/Bis F monomers around carbon nanotubes, two sets of 30 DEDTA and 30 Bis F are placed around (15,15) circular and (36,36) flattened carbon nanotubes. From here forward, the flCNT and circular CNT are referred to (15,15) circular and (36,36) flattened CNTs, respectively. Each simulation has been carried out for 2 million time steps of each 0.25fs for overall time of 500ps. The NVT ensemble and periodic boundary conditions are applied on the simulation boxes of dimensions $60 \times 60 \times 81 \text{ Å}^3$ and $100 \times 60 \times 81 \text{ Å}^3$, for circular and flattened CNTs systems, respectively. The temperature is set to 500K with Berendsen temperature damping constant of 100 fs. All simulations are performed using ADF program [38]. The aim of these simulations is to study the geometrical differences and polymer generation during the simulation. Since we are applying accelerated MD to observe the reactions, the dimensions are selected to make sure there is enough room for all monomers to sit on the carbon nanotubes and not stack above each other.

Next, we investigated the alignment of polymers with the nanotube surface during the simulation. Thus, we developed a script to calculate the orientation of DEDTA molecule, since it has a planar structure. Therefore, three nonadjacent carbon atoms in the aromatic ring of DEDTA molecule are picked and the cross product of the two vectors generated using the 3 carbon atoms positions is considered as the normal vector to the aromatic carbon ring plane. Since the DEDTA molecule is a planar, the normal vector of the aromatic ring can be assumed as the normal vector of the whole molecule. To calculate the alignment of the polymers regarding the surface of the nanotube, the vector which starts from the axis of the circular nanotube to the center of mass of each aromatic ring in DEDTA molecule is considered as the normal vector to the nanotubes surface. Having these two sets of vectors (the normal vectors to the DEDTA plane and surface of nanotube) one can call the Herman orientation function [39] to calculate the alignment of the polymers in relation to the nanotube surface. Equations 4 and 5 represent the Herman function calculations:

$$f = \frac{3(\langle \cos^2 \varphi \rangle - 1)}{2} \tag{4}$$

$$\langle \cos^2 \varphi \rangle = \frac{\int_0^\pi \cos^2 \varphi \sin \varphi \, d\varphi}{\int_0^\pi \sin \varphi \, d\varphi} \tag{5}$$

where $\varphi$ is the angle between the normal vector of DEDTA and nanotube surface and the sign $\langle \ \rangle$ is the average value of variable. If $f = 1$, the maximum, and $f = 0$, random alignments are existing in the system. For the case $f = -0.5$, the vectors are perpendicular to each other.

## 3. Results and discussion

### 3. 1 Polymerization of DEDTA/Bis F in presence of nanotube

DEDTA and Bis F molecules polymerization reaction involves ring opening in epoxy functional group in Bis F and proton transfer from DEDTA to Bis F. Implementing the accelerated ReaxFF simulation, we assured that the exact process will be observed during the simulation since the activation of accelerated MD is designed to be unique, avoiding improper external manipulation to activate other reactions. In other words, the acceleration technique will only focus on the specific reaction between amine and epoxy functional groups and the rest of the system is free to react if required conditions are met. As mentioned before, a system consisting of 30 DEDTA, 30 Bis F and circular CNT was placed in periodic simulation box.

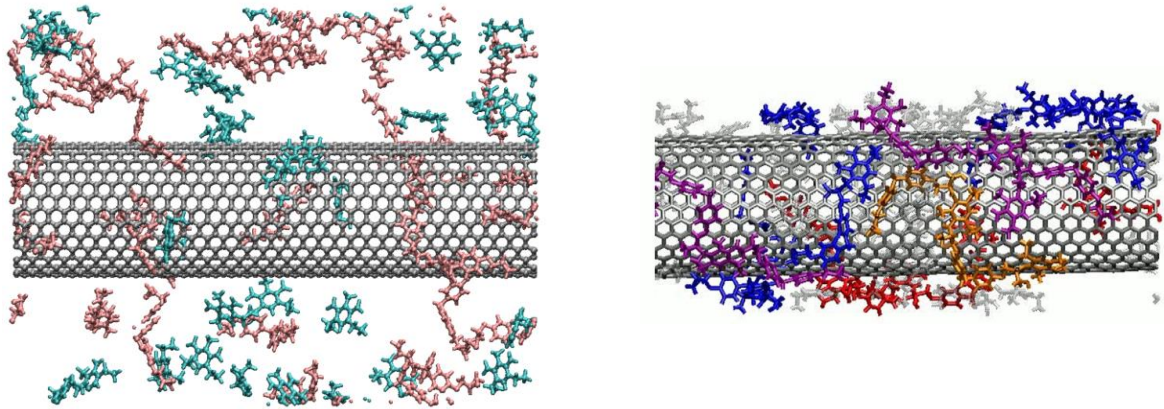

**Figure 9.** (Left) the initial structure of the simulation with 30 DEDTA molecules in cyan and 30 Bis F molecules in pink. (Right) a snapshot from the simulation where the monomers are attached to each other and building dimers, trimers, quadramers and pentamers indicated in blue, red, orange and purple colors, respectively.

Figure 9 (left) illustrates the initial structure of the simulation. The DEDTA and Bis F molecules are represented in cyan and pink colors, respectively. Figure 9 (right) depicts a snapshot during the simulation where the dimers, trimers, quadramers and pentamers are colored by blue, red, orange and purple, respectively. The generated oligomers wrapped around the nanotube and meanwhile, they react with each other, building heavier oligomers. Based on this simulation we observed that the interactions between polymers and nanotube is so strong that the polymers are able to coat the nanotube and not stack on each other. The dominant interaction between the nanotube and polymers are rooted in π-π stacking effect between the aromatic carbon rings in polymers and nanotube surface.

Figure 10 (a) illustrates the initial structure of the system consisting of 30 DEDTA and 30 Bis F in cyan and pink, respectively. Figure 10 (b) is the side view of a snapshot in simulation where heavier polymers are built. It is observed during the simulation that the polymers are moving from the curved region of the flCNT to the flat region. Therefore, the flat region of the flCNT is energetically more favorable than curved region for the polymers to settle and they acquire lower energy value. Based on the binding energy calculations in previous sections, it is shown that the binding energy of DEDTA with flat region of the flCNT is +9.42 kcal/mol stronger than the curved region. The greater curvature in the graphene-like structure of the nanotube leads to lower binding energy value with the adjacent molecule. Thus, this can explain why the generated polymers tend to settle in the flat region of the nanotube.

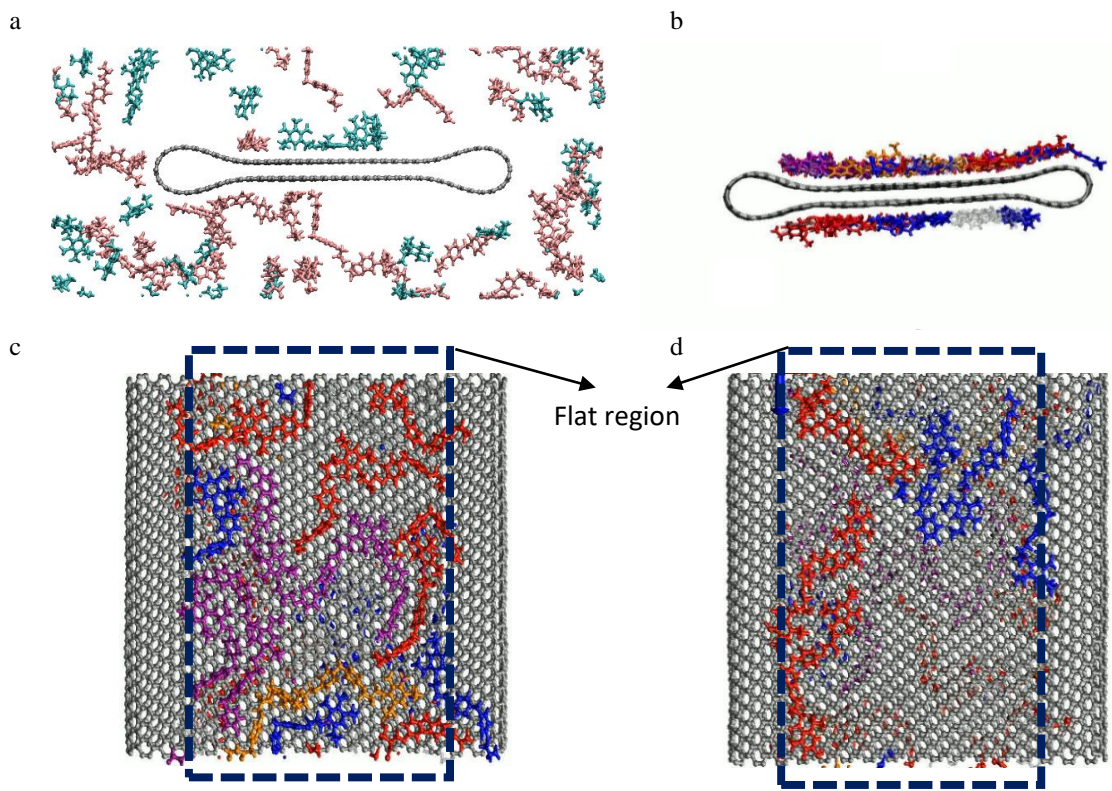

**Figure 10.** (a) Initial structure of the simulation with 30 DEDTA in cyan and 30 Bis F in pink colors. (b) Side view of the snapshot during the simulation with dimers, trimers, quadramers and pentamers in blue, red, orange and purple colors, respectively. (c) and (d), top and bottom view of the flattened carbon nanotube and polymer system with colors as mentioned in (b).

In order to study the rate of reactions and the molecular fractions during the simulations, the number of molecules with different structures are plotted in Figure 11 for systems with circular (left) and flattened (right) nanotubes. For the system with circular nanotube, DEDTA and Bis F are not being consumed evenly. This means that heavier polymers are being built and the transition from small polymers to larger polymers is rapid. On the other hand, for the system with flattened nanotube, DEDTA and Bis F are reacting with each other almost with the same rate, thus, more dimers are generated during the simulation. Comparing the number of the dimer molecules in both simulations, there is 25%-37.5% increase in dimers generation in system with flattened nanotube than that of the system with circular nanotube.

One of the main issues with carbon nanotube-composite systems is the limited dispersion of the nanotubes/polymers causing unbalanced distribution, which can cause stress concentration in composite material. Therefore, higher mobility in polymer molecules leads to higher homogeneity in composite material. Taking in to account that the size of the a polymer has a reverse relation to its mobility, we can conclude that the system with the flattened carbon nanotube has higher chance of homogenous distribution of polymer molecules around the nanotube since the transition from light polymers to heavy polymers are slower than the system with circular nanotube. Therefore, the dimers with higher mobility can disperse in a more effective way than other heavier polymers. This in fact can help manufacturing stronger carbon nanotube-composite UHSL materials, with stronger interactions between the matrix and nanotube along with a more homogenous distribution preventing stress concentrations.

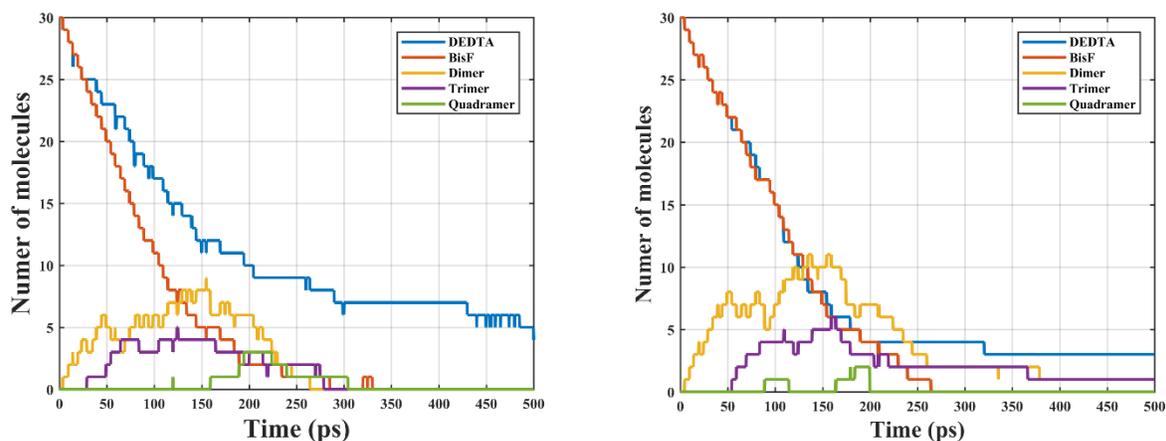

**Figure 11.** No. of molecules consumed and generated during the simulation for the systems with circular nanotube (left) and flattened nanotube (right).

### 3. 2 Alignment of the polymers

In the following sections, we focus on the circular nanotube system to study the alignments of the monomers and the generated polymers in relation to the surface of the nanotube. As discussed before, we took the normal vector of the plane of the aromatic ring in DEDTA as the representative vector of the DEDTA molecule, since it has planar structure. The second vector is the normal vector to the nanotube surface. Figure 12 depicts the normal vectors of DEDTA and nanotube surface in the final frame at t = 500ps in 3D (left) and top (right) views. According to Figure 12, the DEDTA molecules have strong π-π interactions with aromatic rings on the nanotube such that most of them are almost aligned with the surface.

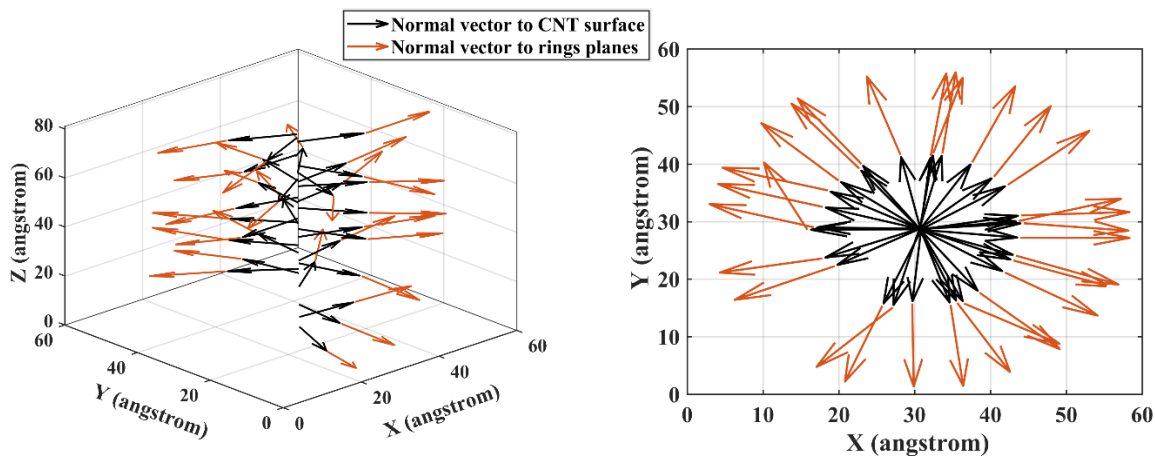

**Figure 12.** 3D (left) and top (right) views of the normal vectors to the aromatic rings of DEDTA and nanotube surface, indicated in red and black colors, respectively. Due to π-π interactions, DEDTA molecules possess high alignments with the nanotube surface.

The vector calculations are performed during the simulation on all frames, therefore, in order to quantitatively investigate the alignment of polymers over nanotube surface, Herman orientation functions (equations 4 and 5) are called. Figure 13 depicts the average angle between the normal vectors (left) and Herman function value (right) during the simulation. At the beginning of the simulation, the molecules are randomly distributed, thus, the Herman value is 0. In the first 100ps of the simulation, all of the monomers are aggregating on the nanotube surface while some of them are being consumed and heavier polymers are being generated. The rapid aggregation of the monomers on the nanotube surface lead to moderately fast increase in Herman value and fast decrease in the average angle between the vectors, in the first 100ps of simulation. Eventually, after 500ps of simulation, the polymers are aligned with the surface with average angle between the normal vectors of <20 degrees and high Herman value of 0.9. We can conclude that based on this simulation, the DEDTA/Bis F polymer matrix can make a highly aligned configuration with the nanotube surface, while the polymerization reaction is taking place.

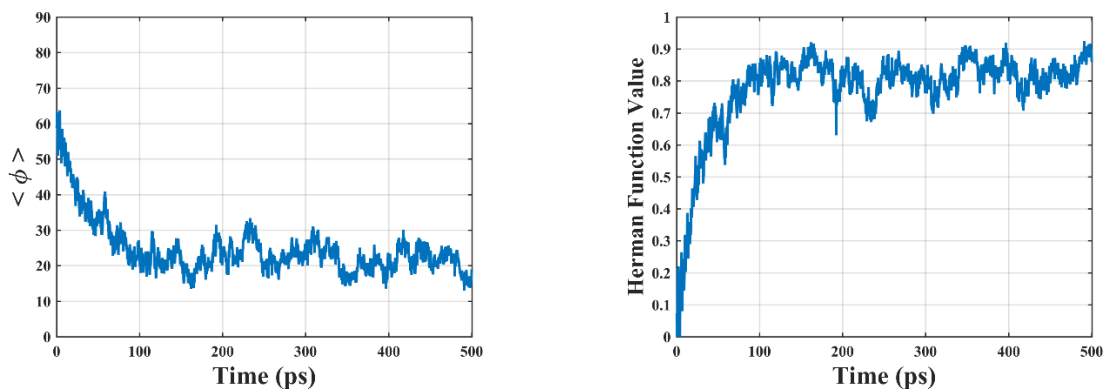

**Figure 13.** Average angle between the normal vectors of nanotube surface and DEDTA plane (left) and the Herman function value (right) during the 500ps of simulation. Based on the diagrams, starting from a random orientation distribution of molecules at t=0, the system obtained high alignment after 500ps.

### 3. 3 Crystallinity of the polymer matrix

One of the important properties of the nanotube-composite material is the crystallinity of the polymer matrix around the nanotube. The crystallinity of a material is defined as a process associated with partial alignment of it which folds, moves and bends together in an ordered pattern. A simulation geometry consisted of 300 BIS F, 300 DEDTA and a circular (15,15) nanotube is generated and placed in a computational box. The system goes through a volume reduction simulation, reducing the size of the box to reach a density of $1 kg/dm^3$. A similar accelerated protocol is performed in 2 million iterations, and once it activates, it will affect the system for 20000 iterations. The temperature of the system is set to 500K with time step of 0.25fs. The simulation box dimensions are $59.2 \times 59.2 \times 81.05$ Å$^3$. Figure 14 (a) illustrates the initial condition of the simulation.

In order to find the layers of crystallinity, we implemented the similar analysis applied for the Herman function calculation and the variations of the angle between DEDTA molecules and nanotube surface are summed up. The cumulative angle variation of the vectors can indicate whether the crystalized layers are formed around the nanotube. Figure 14 (b) depicts the sorted cumulative angle variations of DEDTA molecules in the time span of 250ps to 500ps. This time span is selected to avoid the early stage harsh angle variations of molecules, so they have enough time to settle around the nanotube. Based on the diagram there are two jumps in the angle variations at 5800 and 8000 degrees. Therefore, the area in the diagram is divided to 4 regions. The location of these 4 divisions are plotted in Figure 14 (c). In the first region, colored in black, are the molecules' vectors with cumulative variations < 5800 degrees. As it is illustrated in the Figure 14 (c), these molecules have direct contact with the nanotube surface. The blue colors are for the cumulative angle variations > 8000 degrees. These vectors are mostly located at the edge of the simulation box and the places where the local density is low, thus, they have more freedom to fluctuate. The region between 5800 to 8000 degrees are indicated in orange and red colors. The reason for such differentiation is to see if the red vectors are located closer to the nanotube than orange vectors. According to the Figure 15 (c), there is no significant difference in locations of red and orange vectors and they are scattered randomly inside the simulation box. This analysis revealed that the DEDTA molecules which have already landed on the nanotube (black vectors) are unlikely to swap positions with other molecules and they stick to the nanotube due to strong interactions. Moreover, as shown in previous section, they have reasonably high alignment with the nanotube surface, which in fact strengthens the attraction between the nanotube and DEDTA molecules.

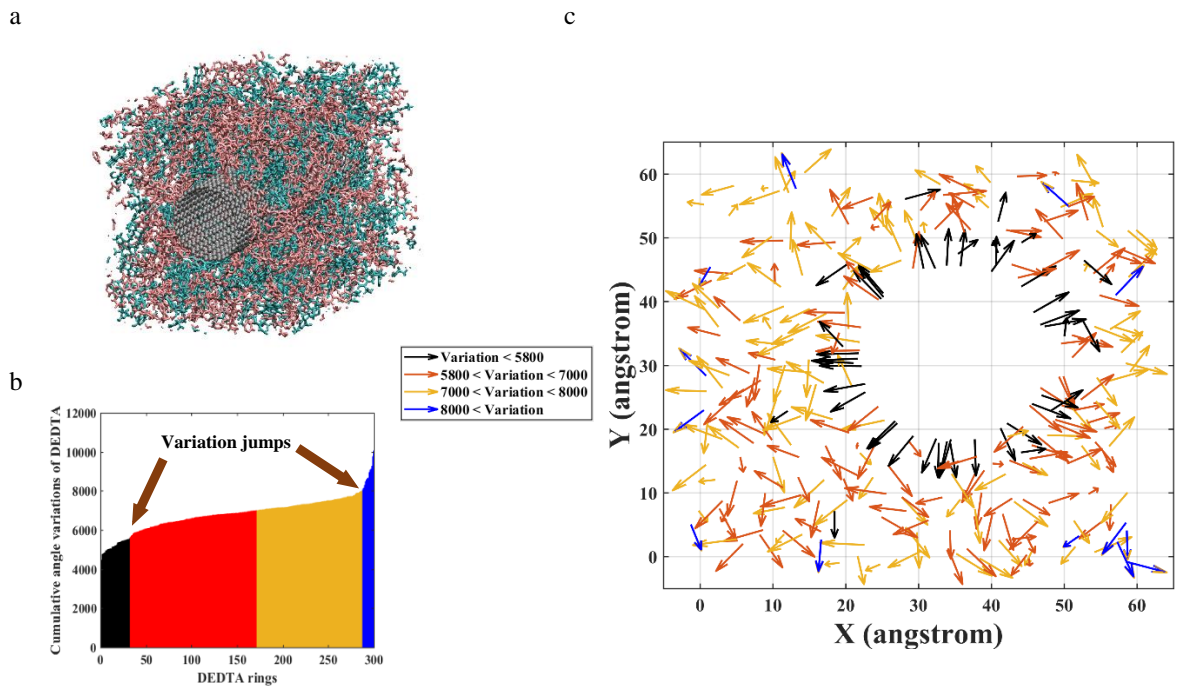

**Figure 14.** (a) Initial condition of the system with 300 DEDTA (cyan), 300 Bis F (pink) and circular nanotube (silver). (b) Cumulative angle variations of DEDTA molecules' normal vectors. (c) Location of the normal vectors of DEDTA molecules with colored based on the cumulative angle variations.

## 4. Conclusion

In this study we optimized the ReaxFF reactive force field carbon parameters to properly capture the shapes and corresponding energy values for flattened carbon nanotubes. We trained the ReaxFF force field carbon parameters against PCFF-IFF force field data which has good agreement to DFT-D2 data. The flattened and collapsed structures of carbon nanotube along with the acquired energy values are given to the ReaxFF training set as input data. Therefore, ReaxFF parameters are optimized for flattened carbon nanotube structures and corresponding energy values. Afterwards, we studied the polymerization of DEDTA and Bis F monomers in vicinity of circular and flattened carbon nanotubes. Our results indicated that in flattened carbon nanotube case, the flat region is energetically more favorable for polymers to settle. Moreover, we observed that more dimers are generated in flattened carbon nanotube – composite system than the circular nanotube – composite system. This observation has a direct impact, providing a more effective coating of nanotube due to high mobility of dimers compared to heavier generated polymers, i.e. trimers, quadramers, etc., leading to higher load transfer between the composite material and flattened nanotube. Moreover, we investigated the alignment of composite materials with respect to the circular nanotube surface. Based on our results, the DEDTA/Bis F polymers set has very high alignment with the nanotube surface due to strong π-π interactions between aromatic carbon rings in the nanotube and polymers. Finally, we studied the layers of crystallinity in the polymer matrix, resulting in the conclusion that the first layers of polymers, adjacent to the nanotube surface, are unlikely to swap positions with the other polymer molecules that are in upper layers.

**Acknowledgement**

We acknowledge funding from NASA/STRI US-COMP.We acknowledge funding from NASA/STRI US-COMP.